\documentclass[12pt]{article}
\usepackage{amssymb,latexsym}            
\usepackage{epsf}                        

\def\be{\begin{equation}}
\def\ee{\end{equation}}
\def\bea{\begin{eqnarray}}
\def\eea{\end{eqnarray}} 
\def\ba{\begin{array}}
\def\ea{\end{array}}

\def\part{\partial}
\def\tfrac#1#2{{\textstyle{#1\over #2}}}
\def\half{\tfrac{1}{2}}
\def\x{\times}
\def\ox{\otimes}

\def\Tr{\mbox{Tr}}

\def\makeatletter{\catcode`\@=11}
\makeatletter
\def\mathbox#1{\hbox{$\m@th#1$}}%
\def\math@ccstyles#1#2#3#4#5#6#7{{\leavevmode
      \setbox0\mathbox{#6#7}%
      \setbox2\mathbox{#4#5}%
      \dimen@ #3%
      \baselineskip\z@\lineskiplimit#1\lineskip\z@
      \vbox{\ialign{##\crcr
             \hfil \kern #2\box2 \hfil\crcr
             \noalign{\kern\dimen@}%
             \hfil\box0\hfil\crcr}}}}
\def\mathaccstyles{\math@ccstyles\maxdimen}
\def\maththroughstyles{\math@ccstyles{-\maxdimen}}
\def\unity%
 {\maththroughstyles{.45\ht0}\z@\displaystyle
 {\mathchar"006C}\displaystyle 1}
\begin{document}

\rightline{IFT-UAM/CSIC-00-39}
\rightline{DTP-00-99}
\rightline{SU-GP-00/{\it 11}-{\it 1}}
\rightline{hep-th/0011242}
\rightline{October 2000}
\vspace{1truecm}

\centerline{\Large \bf Dielectric branes and spontaneous symmetry breaking}
\vspace{1truecm}

\centerline{
    {\bf C\'esar G\'omez${}^{a,}$}\footnote{E-mail address: 
                                  {\tt cesar.gomez@uam.es}}
    {\bf Bert Janssen${}^{a,b,}$}\footnote{E-mail address: 
                                  {\tt bert.janssen@durham.ac.uk} }
    {\bf and} 
    {\bf Pedro J. Silva${}^{a,c,}$}\footnote{E-mail address: 
                                  {\tt psilva@suhep.phy.syr.edu}}}

\vspace{.4truecm}
\centerline{{\it ${}^a$ Instituto de F{\'\i}sica Te\'orica, C-XVI}} 
\centerline{{\it Universidad Aut\'onoma de Madrid}}
\centerline{{\it E-28006 Madrid, Spain}}

\vspace{.4cm}
\centerline{{\it  ${}^b$ Department of Mathematical Sciences}}
\centerline{{\it South Road, Durham DH1 3LE }}
\centerline{{\it  United Kingdom}}

\vspace{.4truecm}
\centerline{{\it ${}^c$ Physics Department, Syracuse University}}
\centerline{{\it Syracuse, New York, 13244, United States}}
\vspace{2truecm}

\centerline{\bf ABSTRACT}
\vspace{.5truecm}

\noindent
A stable non-commutative solution with symmetry breaking is presented
for a 
system of D$p$-branes in the presence of a RR $(p+5)$-form.

\newpage

\noindent
In Type IIA string theory, 
a system of $N$ coinciding D0-branes in the presence of an external RR 
three-form expands in the form of a non-commutative two-sphere
\cite{Myers}
(see also \cite{alekseev}) . This is due to the coupling in the 
(non-commutative) Wess-Zumino term of the D0-brane system to RR forms
of 
degree 3. This result can be easily
generalised to systems of coinciding  $p$-branes in the presence of RR 
$(p+3)$-form.\footnote{For instance, D($-1$) instantons in the presence
of 
RR 2-form expand in a $S^{2}$ manifold and the corresponding instanton
effect can be interpreted as a world sheet instanton wrapping the 
non-commutative manifold.}

\noindent
One specially interesting case, connected with the gravitational
description
of $N^{*}=1$ gauge theories, are D3-branes in Type IIB coupled to a 
seven-form field strength \cite{polchinski}. Generalisations to other 
non-commutative manifolds, like 
$S^{2}\x S^{2}$ or $CP^{2}$, have been recently considered in
\cite{indios}.

\noindent
Another interesting case to study is the system of $p$-branes in the
presence
of a RR $(p+5)$-form. As an example we can have in mind D($-1$)
instantons
in the presence of RR four-form, a physical system that could be
relevant
in the study of instanton effects \cite{green} in the $AdS_{5}$
description 
of $N=4$ supersym\-me\-tric Yang-Mills gauge theory \cite{maldacena}.
In this 
short note we consider this special case, namely D-instantons in
presence
of a RR five-form. The natural guess for a solution would be to look
for some
non-commutative generalisation of the four-sphere. However this
solution can 
easily be shown to be unstable \cite{indios}. Next we will present a
very 
special matrix solution that is a local minimum. Geometrically it can
be 
interpreted as a sort of ellipsoid.

\noindent
The action for D($-1$)-branes in presence of five-form field strength 
(four-form gauge field) is given by
\be
V(X)= \tau \lambda^2 \ \Tr  \Bigl \{ 
          \tfrac{1}{4} [ X_\mu, X_\nu] [ X_\nu, X_\mu] 
        - \tfrac{1}{10} \lambda X^{\sigma\lambda\rho\nu\mu} 
                   F_{\mu\nu\rho\lambda\sigma} \Bigr\} ,
\label{potential}
\ee
where $X^\mu$ are $N \x N$ matrices that represent the
(non-commutative) 
coordinates, $\tau$ the tension of the D-instantons and $\lambda$ the 
string coupling constant. The equation of motion of the above action is
given by
\be
[ [X_\mu, X_\nu], X_\nu] \ - \  
     \half \lambda \ X^{\sigma\lambda\rho\nu}
F_{\mu\nu\rho\lambda\sigma} = 0.
\label{eqnmov}
\ee
To solve this equation we use the Ansatz
\be
X_a = \half R \  (\sigma_a\ox \Sigma_3) \ , \hspace{1cm}
X_i = i \alpha R \ (\unity \ox \Sigma_i)
\label{ansatz}
\ee
where $a=1,2,3$ y $i = 4,5$ and $\vec{\Sigma}$ and $\vec{\sigma}$ are 
generators of $SU(2)$. (We suppose that the matrices $X^A$, with
$A=6...10$, 
commute with the above and amongst each other). These $X^\mu$ satisfy
the following
commutation relations:
\be
\begin{array}{l}
[ X_a, X_b]=\half\ i R^2\ \epsilon_{abc}\  (\sigma_c \ox \Sigma_3^2),
\\
\\

[X_a, X_i]= - \alpha R^2  \ \epsilon_{ij}\ (\sigma_a \ox \Sigma_j) , \\
\\

[ X_i, X_j ] = -2\ i \alpha^2 R^2 \ \epsilon_{ij}\ (\unity \ox
\Sigma_3)   \ .
\end{array}
\label{commrel}
\ee 
Note that these $X^\mu$ do not form an algebra since the commutation
relations
(\ref{commrel}) do not close. Therefore we will obtain a particular
solution, 
rather than a family of solutions.

\noindent
Filling in the Ansatz (\ref{ansatz}) in the equations of motion
(\ref{eqnmov}),
we find for the $a$ and $i$-component respectively:
\be
\ba{l}
R^3 \Bigl[1- 4\alpha^2 -3\lambda f \alpha^2 R \Bigr] \
                  (\sigma_a \ox \Sigma_3) = 0 \ , \\ 
\\
i \alpha R^3  \Bigl[ 3- 4\alpha^2 + \tfrac{3}{2}\lambda f  R \Bigr]
                   \ (\unity \ox \Sigma_i) = 0\ .
\ea
\label{eqn}
\ee
Here we had to suppose that $\Sigma_i^2 = \sigma_a^2 = \unity$, which
implies
that 
$\vec{\Sigma}$ and $\vec{\sigma}$ are either two-dimensional
(irreducible) 
representations of $SU(2)$ or $4n$-dimensional reducible
representations,
where $n=n_\sigma n_\Sigma$ and $n_\sigma$ ($n_\Sigma$) is the number
of 
irreps in the reducible representation of $\vec\sigma$
($\vec{\Sigma}$). 
Thus we can only describe systems with a number of D($-1$)-branes which
is a 
multiple of four.

\noindent
Clearly, the equation of motion (\ref{eqn}) are satisfied for a
(trivial) 
commutative solution $\alpha=R=0$, but also a non-trivial
non-commutative one 
given by
\be
R= \frac{1-4\alpha^2}{3 \lambda f \alpha^2}  \hspace{1cm}
\mbox{and}    \hspace{1cm}
R= \frac{8\alpha^2 - 6}{3 \lambda f} \ ,
\ee
which leads to two solutions
\bea
&& \alpha_+^2 = \frac{1}{2}, \hspace{3cm} R_+ = \frac{-2}{3\lambda f} \
; 
\label{solutions1}\\
&& \alpha_-^2 = -\frac{1}{4}, \hspace{2.7cm} R_- = \frac{-8}{3\lambda
f} \ . 
\label{solutions2}
\eea
Notice that due to the non-trivial value of $\alpha$, this solutions
can be 
interpreted as a sort of ellipsoid.

\noindent 
To study the stability of these solutions we use the Ansatz
(\ref{ansatz}) 
in the potential (\ref{potential}), 
\be
V(\alpha, R) = \lambda^2 \tau R^4 n 
  \Bigl[ \tfrac{3}{2} - 12 \alpha^2 + 8 \alpha^4 - 6\lambda f\alpha^2 R
\Bigr],
\label{potential2}
\ee
which evaluated in the solutions (\ref{solutions1})-(\ref{solutions2})
gives 
\be
\ba{l}
V_+ =-\frac{1}{4} \lambda^2 \tau R^4 n \ ,\\ \\
V_- = \lambda^2 \tau R^4 n \ .
\ea
\ee
Thus we see that the solution $(\alpha_+, R_+)$ has lower energy than
the 
commutative solution, while the solution $(\alpha_-, R_-)$ has higher
energy.
Calculating the second variation of the potential (\ref{potential2}),
it is 
easy to see that the solution $(\alpha_+, R_+)$ corresponds to a
minimum,
while $(\alpha_-, R_-)$ is a saddle point. 

\noindent
Therefore we can conclude
that, in 
the presence of a five-form field strength, a set of $4n$ D-instantons
will
decay spontaneously into the stable solution $(\alpha_+,
R_+)$.\footnote{In
\cite{Myers} is was observed that reducible $N\x N$ representations
have 
higher (though still negative) energies than the irreducible $N\x N$ 
representations. However, as we noticed above, our solution for general
$N$, 
only exists  for reducible representations.}

\noindent
Notice that this solution breaks $SO(5)$ invariance of the potential 
(\ref{potential}). However we observe that
the solution is degenerated with respect to 
permutations of the $\Sigma$ matrices: any set of matrices of the form
\be
X_a = \half R \  (\sigma_a\ox \Sigma_{j_0}) \ , \hspace{1cm}
X_i = i \alpha R \ (\unity \ox \Sigma_i)
\label{ansatz2}
\ee
with $i$ different from $j_0$, is also a solution with identical
behaviour as
 (\ref{solutions1})-(\ref{solutions2}). Thus this solution could be
interpreted as 
a sort of spontaneous symmetry breaking for the matrix action. The
extension
to any $(p, p+6)$ system is straight forward.

\noindent
As a spinoff of our analysis we just mention that this solution can be 
naturally used to construct longitudinal five-brane solutions in
M(atrix)
theory (see for instance \cite{rand} and references therein).

\vspace{1cm}
\noindent
{\bf Acknowledgements}\\
We wish to thank Patrick Meessen and Enrique \'Alvarez 
for the useful discussions.
The work of C.G. and B.J. has been supported by the TMR program
FMRX-CT96-0012
on {\sl Integrability, non-perturbative effects, and symmetry in
quantum field
theory}. The work of P.J.S. has been supported in part by NSF grant
PHY97-22362 
to Syracuse University and by funds from Syracuse University.


\end{document}